\documentclass{article}
\usepackage[utf8]{inputenc}
\usepackage[english]{babel}
\usepackage[a4paper,top=2cm,bottom=2cm,left=3cm,right=3cm,marginparwidth=1.75cm]{geometry}
\usepackage{xcolor}
\usepackage{amsmath}
\usepackage{graphicx}
\usepackage{authblk}

\title{Effects of tinted lenses on chromatic sensitivity:  changes in colour vision assessed with the CAD test, a preliminary study}

\author[1]{Lucia Natali}
\author[1,2,3]{Alessandro Farini}
\author[2,3]{Elisabetta Baldanzi}
\author[4]{John Barbur}
\affil[1]{IRSOO Istituto Ricerca e Studi in Ottica e Optometria, Vinci, Italy}
\affil[2]{Istituto nazionale di Ottica-CNR, Largo Fermi 6, 50125 Firenze, Italy}
\affil[3]{Università di Firenze, corso di laurea in Ottica e Optometria, Firenze, Italy}
\affil[4]{Centre for Applied Vision Research, School of Health Sciences, City, University of London, London, UK}

\date{\today}

\begin{document}

\maketitle

\begin{abstract}
The aim of this study was to assess the extreme effects of tinted lenses on colour vision by examining changes in chromatic sensitivity when viewing visual displays through a slightly tinted (``blue-blocking'') filter and through a heavily tinted, ``orange'' coloured filter.  
The CAD test was used to measure both red/green (RG) and yellow/blue (YB) chromatic sensitivity in ten subjects when viewing visual displays through each of the two filters. 
The measured RG and YB colour thresholds were then compared with similar measurements made without coloured filters in front of the eye. 
The blue-blocking filter absorbs only a small amount of short-wavelength light whilst the ``orange'' filter attenuates preferentially more short wavelength and some middle-wavelength light. 
The results show that the blue-blocking filter does not affect significantly either RG or YB colour vision. 
The orange filter, on the other hand, causes large changes in colour discrimination. 
The results were analysed statistically by comparing results obtained with the coloured filters with those measured without any filters in front of the eye. 
More experimental work is now needed to establish how much short wavelength light can be removed without affecting significantly the subject’s colour discrimination performance.   
\end{abstract}

\section{Introduction}
Colour vision is important since colour signals can be used to code information that improves our visual performance. The ability to discriminate different colours has many advantages in daily and professional life. 
Furthermore, it is mandatory to confirm compliance with safety standards in visually-demanding working environments where good colour perception is required (aviation, maritime transport, fire and rescue services, etc)\cite{barbur2017colour}. 

People use lenses during their daily activities, because they need a correction for their refractive conditions or because they want to reduce the amount of light reaching the retina and also to minimise glare. 
Since the use of various types of ophthalmic lenses that are often spectrally selective is widespread, it is of interest to understand how the use of such lenses can affect colour vision. 
In particular, we would like to predict accurately how the subject’s colour discrimination abilities are affected by the use of coloured lenses and to understand the possible repercussions the use of coloured lenses may have in normal daily life and in working environments. 
The immediate objective of this work is to evaluate how colour discrimination on visual displays varies when using coloured lenses in front of the eye.  
We report preliminary results which measure red/green (RG) and yellow/blue (YB) colour thresholds when viewing visual displays through coloured filters.

\section{Tinted lens characteristics}
\begin{figure}[htbp]
    \centering
    \includegraphics[width=\linewidth]{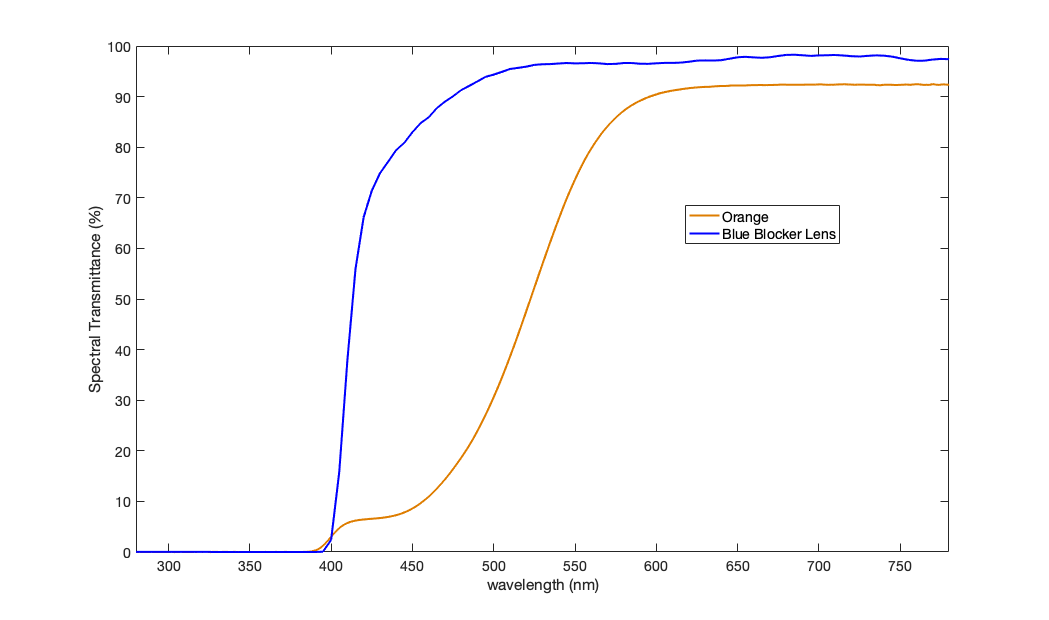}
    \caption{Spectral transmittance of the two lenses used in the experiment. The blue line is the blue blocker lens (Essilor 
    Crizal Stylis, $n=1.67$), the orange line is an orange lens.}
    \label{fig:spettri}
\end{figure}

Tinted lenses nowadays could be realised using two different methods. 
The traditional method uses the absorption properties of pigments, but in recent years it is common to use bespoke interference filters with selective absorption bands. 
Filters based on thin, multi-layer coatings can be designed to control accurately the amount of short-wavelength light absorbed by the lens. 
These two methods of producing coloured filters affect both the amount and the spectral composition of the transmitted light. 
Recently blue blocking lenses based on thin-layer coatings have been produced and made available commercially by lens manufacturers, claiming good protection from blue light\cite{comparetto2019mitigating}. 
In our experiment we use two different lenses, one that relies on light absorption by spectrally-selective pigments, the other one a blue blocker lens produced multilayer thin film interference (Fig.\ref{fig:spettri}). 
Transmittance measurements were made using a Perkin Elmer 1050 spectrophotometer.

\section{Subjects}
Ten subjects (six men and four women), who ranged in age from 19 to 26 (mean age$=22.1\pm2.4$), participating in the study. 
We prefer to limit the age interval for our subjects because we want to study only the effect of tinted lenses, avoiding well-known effects on colour vision related to age\cite{barbur2016color}.
The subjects had binocular visual acuity better than one minute of arc (i.e. equivalent to 20/20 on the Snellen chart) and needed no refractive corrections to see clearly the test stimuli on the visual display.

\section{Colour test}
\begin{figure}
    \centering
    \includegraphics{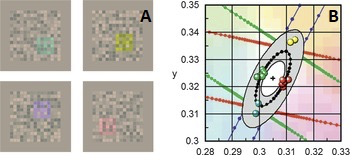}
    \caption{A) RG and YB stimuli employed in the CAD test. B) The statistical limits for the standard normal (SN) CAD observer are plotted in the CIE(x,y) 1931 chromaticity chart.}
    \label{fig:cad}
\end{figure}

We use the Colour Assessment and Diagnosis (CAD) test in our experiment. The CAD test isolates the use of coloured signals and measures colour detection thresholds. 
Studies on camouflage showed that the use of dynamic luminance contrast noise masks the detection of luminance contrast signals without affecting significantly either RG or YB chromatic sensitivity\cite{barbur2004double}. 
The test involves the use of a calibrated monitor, keyboard and a computer on which CAD software is installed. 
The test generates coloured stimuli on a visual display within a background of dynamic luminance contrast noise. 
The stimuli move along each diagonal direction, and the colours are selected to ensure automatic classification of the class of deficiency involved, as well as adequate estimates of both RG and YB colour thresholds (Fig.\ref{fig:cad}A).

The subject’s task is to indicate the direction of movement of the colour-defined stimulus using one of the four, raised, corner buttons on the numeric keypad. 
The test requires the subject’s age which is an important parameter since the results are compared against the expected normal age limits. 
When the full test is carried out the results show the subject’s RG and YB thresholds in CAD units. 
A graph is also displayed to indicate the measured RG and YB thresholds superimposed on the normal, age-matched chromatic threshold ellipse. 
The latter is plotted in the CIE (x,y)-1931 chromaticity diagram, together with the expected upper and lower threshold limits for the subject’s age (Fig.\ref{fig:cad}B).

\section{Results}
\begin{figure}
    \centering
    \includegraphics[scale=0.4]{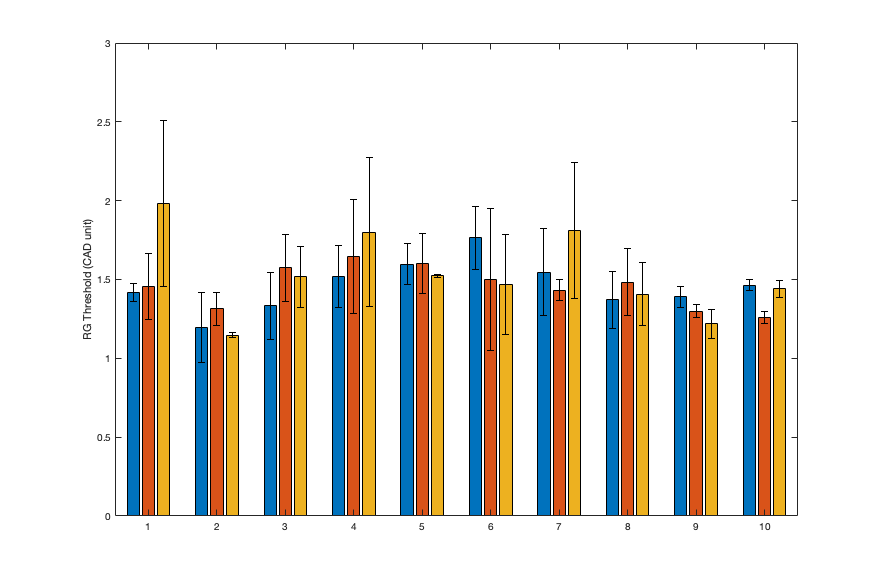}
    \caption{Mean RG thresholds measured in the ten study participants for each of the three viewing conditions: direct viewing without any lens in front of the eye (blue bars), with the blue-blocker lens (red bars) and with the orange lens (orange bars). The vertical lines indicate the computed inter-subject variability $(\pm\sigma)$}
    \label{fig:soglieRG}
\end{figure}

\begin{figure}
    \centering
    \includegraphics[scale=0.6]{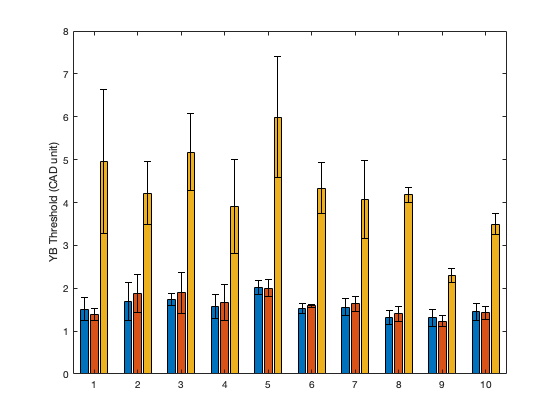}
    \caption{Mean YB thresholds measured in the ten study participants for each of the three viewing conditions: direct viewing without any lens in front of the eye (blue bars), with the blue-blocker lens (red bars) and with the orange lens (orange bars). The vertical lines indicate the computed inter-subject variability $(\pm\sigma)$}
    \label{fig:soglieYB}
\end{figure}

The thresholds for the RG (Fig.\ref{fig:soglieRG}) and the YB (Fig.\ref{fig:soglieYB}) channels show that the orange lens causes a large loss of colour vision, while the blue blocker lens does not affect significantly either RG or YB colour thresholds.

\begin{figure}[htbp]
    \centering
    \includegraphics[scale=0.6]{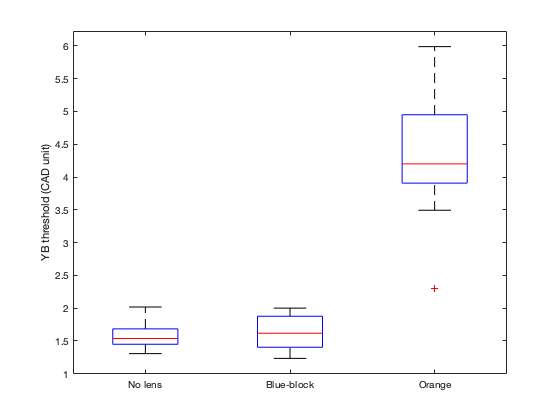}
    \caption{Box plot for YB threshold. The red line represents the median, the whiskers are the maximum and minimum value, the boxes are the 25 and 75 percentiles: the plot shows the large difference between the orange lens and the other two conditions.}
    \label{fig:boxPlot}
\end{figure}

In addition, we carried out a one-way ANOVA to compare the effect of type of lens on RG and YB threshold. 
This test revealed that there was a statistically significant difference in YB threshold between at least two groups $(F(2,27)=[65.24]$, $p<10^{-5})$. 
Tukey's HSD test for multiple comparisons revealed significant differences in YB thresholds when comparing the no lens and orange lens conditions ($p<10^{-5}$, $95\% C.I.=[-2.6967,-2.0265]$) and the blue-blocker lens and the orange lens ($p<10^{-5}$, $95\% C.I.=[-2.6503,-1.9802]$). 
The box plot for YB threshold is shown in Fig.\ref{fig:boxPlot}. 
There was no statistically significant difference between no lens and blue-blocker lens conditions ($p=0.98$). 
Regarding RG threshold, the one-way ANOVA revealed no statistically significant differences between the three conditions.

\begin{figure}[htbp]
    \centering
    \includegraphics[scale=0.3]{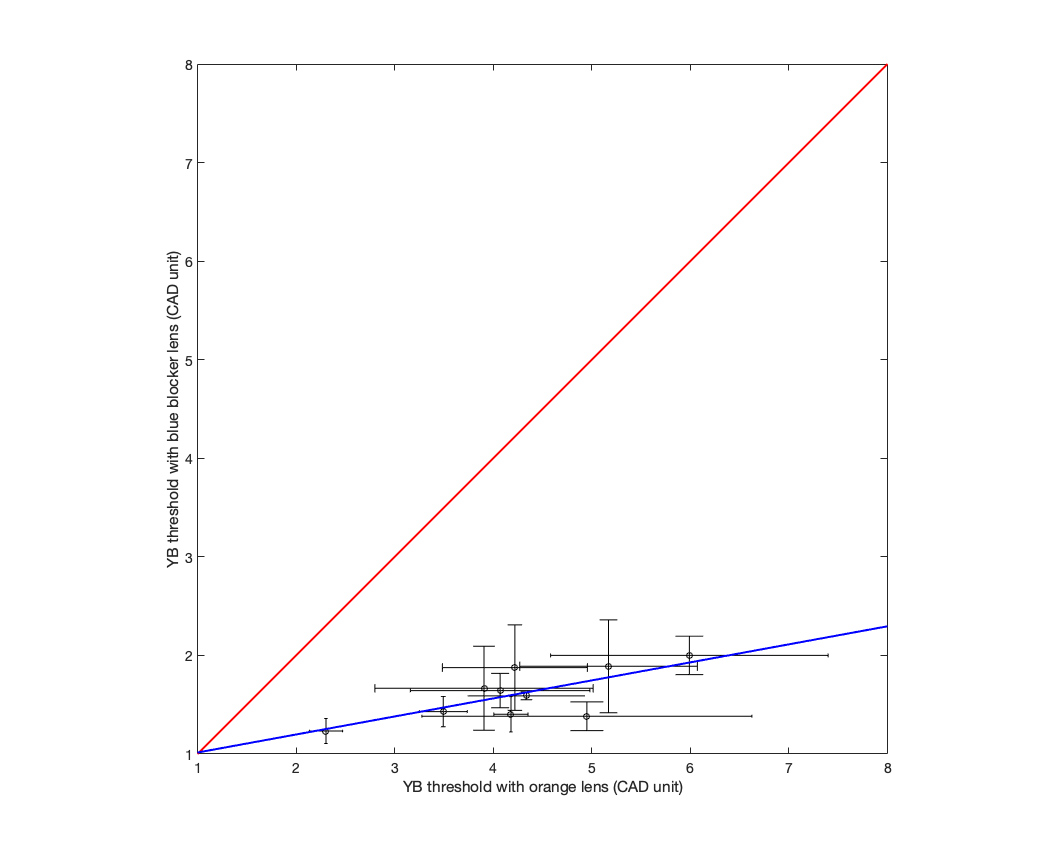}
    \caption{Relationship between YB threshold with orange lens ($x$ axis) and with blue blocker lens ($y$ axis). The error bars are the standard deviations. The blue line is the regression line $(y=0.18x+0.83)$, the red line is the $y=x$ line. Pearson’s $r = 0.72$.}
    \label{fig:orangevsblue}
\end{figure}

\begin{figure}[htbp]
    \centering
    \includegraphics[scale=0.3]{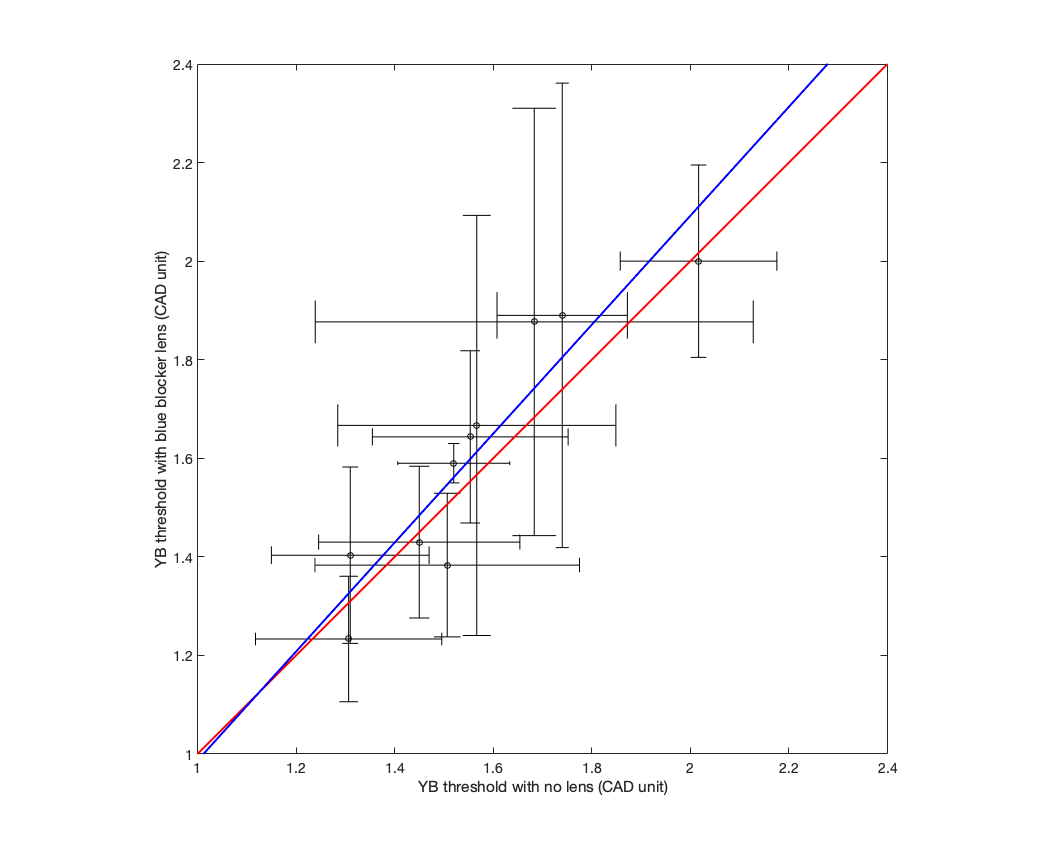}
    \caption{Relationship between YB threshold with no lens ($x$ axis) and with blue blocker lens ($y$ axis). The error bars are the standard deviations. The blue line is the regression line ($y=1.10x-0.11$), the red line is the $y=x$ line. Pearson’s $r=0.92$.}
    \label{fig:novsblue}
\end{figure}

Comparison of results plotted in Fig.~\ref{fig:orangevsblue} and Fig.~\ref{fig:novsblue} confirms that the orange lens strongly affects colour discrimination, whilst the effect of the blue-blocker lens is negligible.

\section{Conclusions}
The findings reported here show that although commercial blue-blocking lenses have a negligible effect on colour discrimination on visual displays, coloured lenses that absorb a greater amount of short wavelength light can cause large loss of chromatic sensitivity with the YB channel being most affected. 
This preliminary study demonstrates that the CAD test could be used to assess objectively the maximum amount of short-wavelength light one can absorb with blue-blocking spectacle lenses without affecting significantly either RG or YB chromatic sensitivity.

\newpage
\bibliography{bibliografia}
\bibliographystyle{unsrt}

\end{document}